\documentclass[pra,reprint,superscriptaddress,nobibnotes,floatfix]{revtex4-1}

\usepackage[T1]{fontenc}
\usepackage[utf8]{inputenc} 

\usepackage{amsmath,amssymb,graphicx,bm,microtype}
\usepackage[dvipsnames]{xcolor}
\usepackage{booktabs}
\usepackage{siunitx}
\usepackage{braket}
\usepackage[colorlinks,allcolors=blue!50!black]{hyperref} 
\usepackage[all]{hypcap}

\newcommand{\ketbra}[2]{\ket{#1}\!\bra{#2}}
\newcommand{\mb}{\mathbf}
\newcommand{\uq}{_{\mathbf{q}}}
\newcommand{\uql}{_{\mathbf{q}\lambda}}

\newcommand\scalemath[2]{\scalebox{#1}{\mbox{\ensuremath{\displaystyle #2}}}}

\begin{document}

\title{Environmentally Improved Coherent Light Harvesting}

\author{Stefano Tomasi}
\affiliation{School of Chemistry and University of Sydney Nano Institute, University of Sydney NSW 2006, Australia}

\author{Dominic M. Rouse}
\affiliation{SUPA, School of Physics and Astronomy, University of St Andrews, St Andrews KY16 9SS, UK}

\author{Erik M. Gauger}
\affiliation{SUPA, Institute of Photonics and Quantum Sciences, Heriot-Watt University, Edinburgh EH14 4AS, UK}

\author{Brendon W. Lovett}
\affiliation{SUPA, School of Physics and Astronomy, University of St Andrews, St Andrews KY16 9SS, UK}

\author{Ivan Kassal}
\email[Email: ]{ivan.kassal@sydney.edu.au}
\affiliation{School of Chemistry and University of Sydney Nano Institute, University of Sydney NSW 2006, Australia}

\begin{abstract}
Coherence-enhanced light harvesting has not been directly observed experimentally, despite theoretical evidence that coherence can significantly enhance light-harvesting performance. The main experimental obstacle has been the difficulty in isolating the effect of coherence in the presence of confounding variables. Recent proposals for externally controlling coherence by manipulating the light's degree of polarization showed that coherent efficiency enhancements would be possible, but were restricted to light-harvesting systems weakly coupled to their environment. Here, we show that increases in system-bath coupling strength can amplify coherent efficiency enhancements, rather than suppress them. This result dramatically broadens the range of systems that could be used to conclusively demonstrate coherence-enhanced light harvesting or to engineer coherent effects into artificial light-harvesting devices.
\end{abstract}

\maketitle

Light-harvesting systems absorb energy from a light source and transport the resulting exciton to an acceptor where it can drive chemical or physical processes such as photosynthesis or photovoltaic current generation.
A key figure of merit for light harvesting is the efficiency (or quantum yield), the probability that excitons reach the acceptor instead of being lost to recombination during their short lifetime.

Theoretical studies have proposed a large number of detailed mechanisms by which excitonic coherence could enhance light-harvesting efficiency~\cite{Cao2009,Dorfman2013,Svidzinsky2011,Scully2010,Scully2011,Creatore2013,Kassal2013,Wu.20139io,Leon-Montiel2014,Olsina2014,Dodin2016,Dodin2016a,Baghbanzadeh2016,Baghbanzadeh2016_2,Oviedo-Casado2016,Fruchtman2016,Higgins2017,Tscherbul2018,Brumer2018,Brown2019,Rouse2019,Tomasi2019,Tomasi2020,Yang.2020,Jankovic2020}. This variety of mechanisms is partly due to the variety of meanings of the word ``coherence''; in particular, the common definition of coherences as the off-diagonal elements of a density matrix makes them basis-dependent concepts~\cite{Kassal2013,Tomasi2020}. Here, unless specified otherwise, we will use the word ``coherence'' to refer to off-diagonal density-matrix elements in the eigenbasis of the system Hamiltonian.

Despite the various theoretical proposals, no experiment has unambiguously demonstrated an efficiency increase in light harvesting due to any type of excitonic coherence.
The difficulty lies in creating an experimental control for comparing efficiencies with and without coherence, without introducing confounding variables such as changes in chemical or physical structure~\cite{Baghbanzadeh2016_2}.
Fortunately, coherence between a system's excitonic states can be controlled externally by changing the characteristics of the exciting light source, whether spectral coherence~\cite{Jiang1991,Bruggemann2004,Voronine2011,Caruso2012,Hoyer2014,Mancal2010} or polarization~\cite{Tscherbul2018,Tomasi2019,Yang.2020}.

The ways in which different types of coherence can improve the efficiency have been classified~\cite{Tomasi2020}. Whether an enhancement occurs depends on the nature of excitation and exciton trapping~\cite{Leon-Montiel2014}; in particular, the efficiency can be increased by coherence in the eigenbasis if environmental influences act in a different basis (or vice versa)~\cite{Tomasi2020}.

\begin{figure*}[tb]
    \centering
    \includegraphics[width=0.8\linewidth]{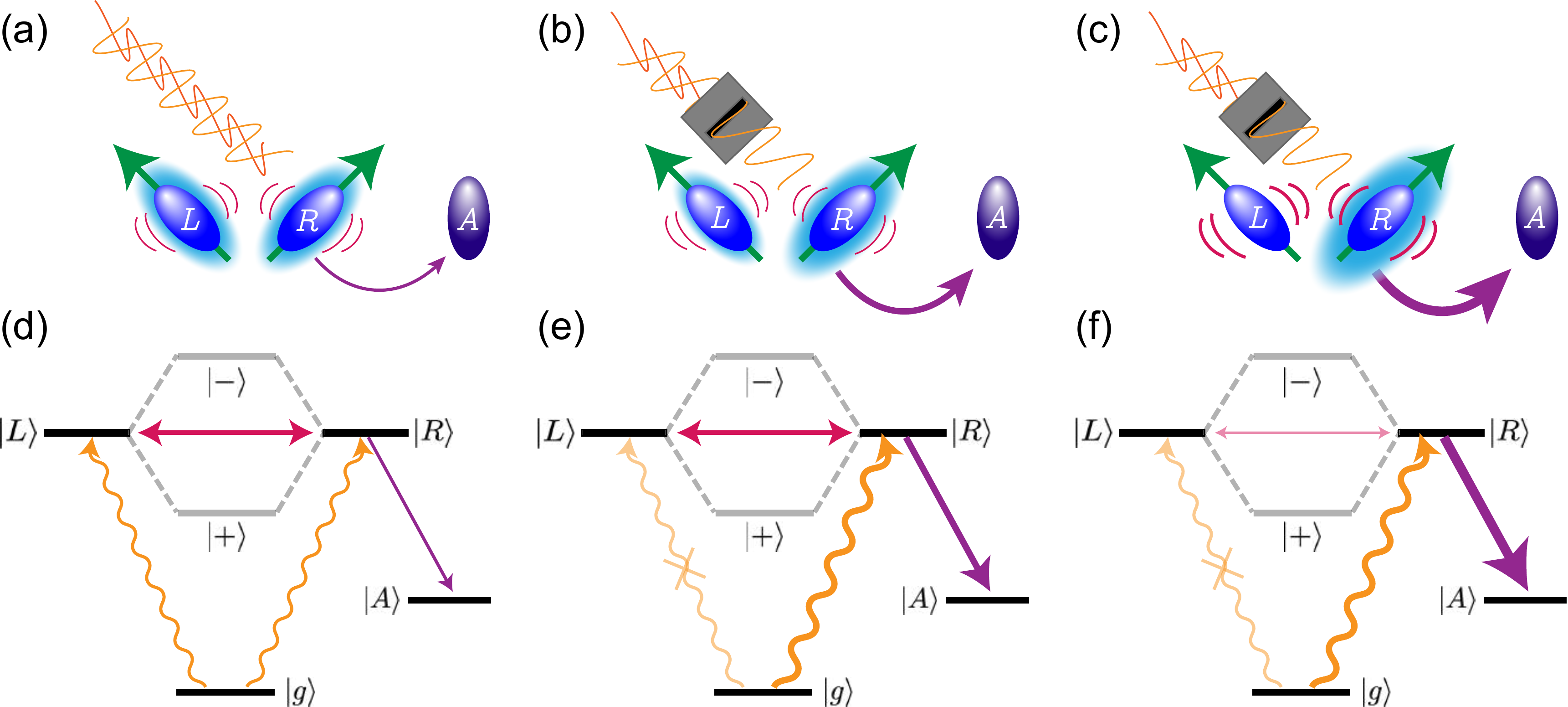}
    \caption{Coherence-enhanced light harvesting amplified by the environment. Our system contains two light-absorbing donors (blue, $L$ and $R$) and an acceptor (purple, $A$) to which excitons can be transferred. Donor transition dipole moments (green arrows) are perpendicular and $A$ is excitonically coupled only to $R$.
    \textbf{(a)} Unpolarized light creates an equal mixture of excitons on both sites, leading to the lowest efficiency because only excitons from $R$ can be trapped directly.
    \textbf{(b)} Polarized light parallel to $R$'s transition dipole gives a higher efficiency by exciting a coherent superposition of eigenstates that is localized on $R$, but some efficiency is lost due to excitons being transferred between donors. (Generating excitons localized on $L$ would result in a decreased efficiency.)
    \textbf{(c)} The same polarized light but with strong system-bath coupling gives the highest efficiency because increased coupling to the environment suppresses exciton transfer between donors, keeping them localized on $R$, where they can be trapped.
    \textbf{(d--e)} The corresponding system states and exciton transfer pathways, including pumping from the ground state to donor sites (orange arrows), transfer to acceptor (purple arrows), transfer between donors (red arrows), and donor eigenstates $\ket{\pm}$ (grey lines).}
    \label{fig1}
\end{figure*}

We recently proposed a specific, minimal light-harvesting system composed of two identical chromophores coupled to a single acceptor, where coherence in the eigenbasis can be externally controlled through the polarization of a light source~\cite{Tomasi2019}, allowing changes in efficiency to be directly attributed to changes in coherence.
The proposed mechanism relies on forming localized states from coherent superpositions of delocalized eigenstates, meaning that coherence can be used to produce an exciton localized close to the acceptor, enhancing transfer rates compared to those from delocalized states.
The model's main limitation is its assumption of weak system-bath coupling, which limits its range of applicability.

Here, we show that, surprisingly, increases in system-bath coupling can amplify the efficiency enhancements caused by coherence.
We do so by extending the model in~\cite{Tomasi2019} using the polaron transformation~\cite{nazir2016modelling,qin2017effects,mccutcheon2011general,pollock2013multi,jesenko2014excitation,Lee2015}, which allows us to describe the light-harvesting efficiency of systems with intermediate or strong system-bath coupling.
Our results imply that a strongly coupled bath can be advantageous both for demonstrating coherence-related efficiency enhancements in model systems and for engineering them into artificial light-harvesting platforms.

Our system, based on ref.~\cite{Tomasi2019} and shown in Figure~\ref{fig1}, comprises a pair of identical light-harvesting sites (donors), excitonically coupled to each other to form a dimer. One of the sites is also coupled to an acceptor, where excitons are to be transferred.
Since only one donor is coupled to the acceptor, the efficiency of the transfer to the acceptor is maximized from excitons that are localized on this site.
Because the excitonic eigenstates of the dimer are delocalized across both sites~\cite{maykuhn}, localized states are coherent superpositions of eigenstates, where constructive interference occurs on one site and destructive interference on the other.
Therefore, in this system, efficiency can be increased by controlling the coherence so that the exciton is intially on the site nearer the acceptor and spends more time there than on the other site~\cite{Tomasi2019}, a Type~IIA enhancement in the classification of~\cite{Tomasi2020}.

Demonstrating that coherence can enhance efficiency requires specific control over the coherence, so that efficiencies with and without it can be compared, and with other parameters held constant.
To achieve this control, our donors are arranged with perpendicular transition dipole moments, so that each donor is excited by different polarization modes of the light.
In this arrangement, light polarized parallel to a donor's dipole generates excitons in a localized, coherent superposition of eigenstates (Figure~\ref{fig1}b), while unpolarized light generates a statistical mixture (Figure~\ref{fig1}a)~\cite{Tscherbul2018,Tomasi2019,Yang.2020}.
If there were no subsequent excitonic dynamics in the dimer, the coherence of the initial states could be used to enhance the efficiency by initialising the exciton close to the acceptor.

However, following excitation, exciton dynamics between the donors can take place, either coherently or driven by coupling to the environment. 
In that case, the efficiency will be enhanced by the initial coherence only if the dynamics is slower than the trapping, ensuring that an exciton localized close to the acceptor remains there long enough to be trapped~\cite{Tomasi2019}.
The simplest way of ensuring slow inter-donor dynamics is with weak donor-donor and system-bath couplings~\cite{Tomasi2019}. 

Here, we relax the assumption of weak couplings to show that coherent efficiency enhancements are possible even with strong system-bath couplings. Indeed, as depicted in Figure~\ref{fig1}c, a stronger system-bath coupling can further amplify the enhancement because it suppresses all dynamics between the donors.

\begin{figure*}[tb]
    \centering
    \includegraphics[width=\textwidth]{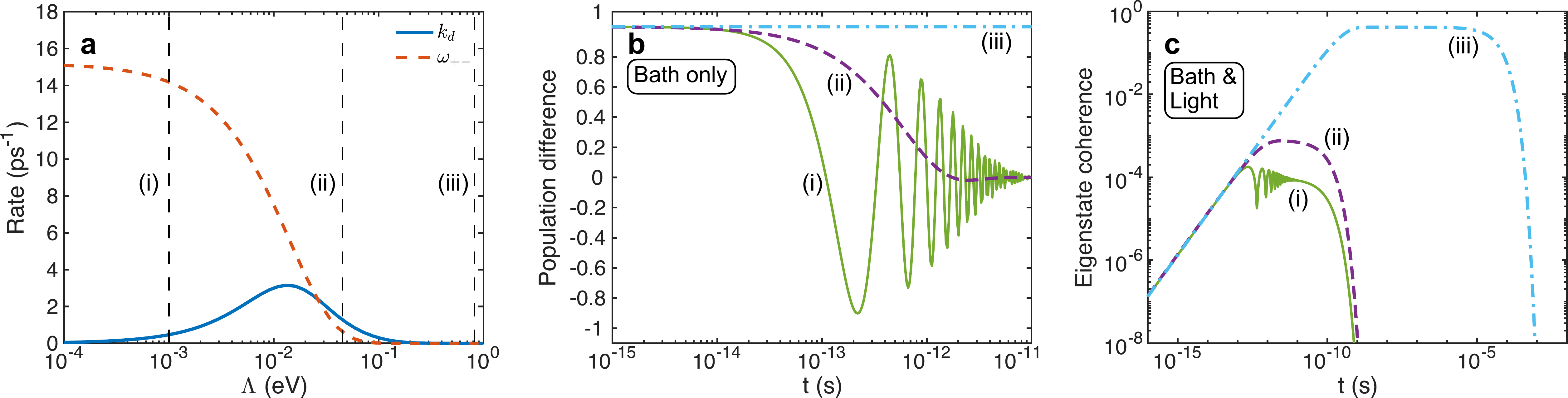}
    \caption{Strong system-bath coupling suppresses exciton transfer between the two donor sites and preserves coherences for longer. In these calculations, no acceptor is present.
    \textbf{(a)} Oscillation frequency $\omega_{+-}$ and dephasing rate $k_d$ (Eq.~\ref{eq:coh_func}) as functions of reorganization energy $\Lambda$. 
    \textbf{(b)} Population difference between the two donor sites caused by the bath interaction for an initially localized state.
    \textbf{(c)} Coherence $\left|\rho_{+-}(t)\right|$ between the eigenstates of the two donors, with coupling to both the bath and the light. In all cases, the system is initialized in the ground state and the light is polarized to pump the coherence according to Eq.~\eqref{eq:cohgen}. 
    In \textbf{(b)} and \textbf{(c)}, the three curves correspond to the three values of $\Lambda$ labeled in \textbf{(a)} and showing three distinct regimes: \textbf{(i)} low system-bath coupling allows underdamped oscillations, \textbf{(ii)} stronger coupling leads to an overdamped regime and \textbf{(iii)} extreme coupling suppresses the dynamics for long times. 
    Once an acceptor is included that traps locally from a single site, the extended localization in \textbf{(iii)} maximizes trapping efficiency. Donor and bath parameters as in Table~\ref{tab:params}, except for variable $\Lambda$. }
    \label{fig2}
\end{figure*}

We start with a Frenkel Hamiltonian for the donor dimer, with each site---whether a molecule, artificial atom, or semiconductor nanostructure---coupled to an independent bath of harmonic oscillators (e.g., nuclear vibrational modes):
\begin{subequations}\label{eq:exciton_ham}
\begin{align}
    H &= H^S + H^{SB} + H^{B} \\
    H^S &= \sum_{s\in\left\{L,R\right\}}\delta_s\ketbra{s}{s}+V(\ketbra{L}{R} + \ketbra{R}{L}) \\
    H^{SB} &= \sum_{s\in\lbrace L,R\rbrace} \sum_\xi g_{s,\xi} \omega_\xi (b^\dag_{s,\xi}+b_{s,\xi})\ketbra{s}{s} \\
    H^B &= \sum_{s\in\lbrace L,R\rbrace} \sum_\xi \omega_\xi b^\dag_{s,\xi}b_{s,\xi},
\end{align}
\end{subequations}
where $\ket{L}$ and $\ket{R}$ are states with excitons localized on the left and right sites respectively, $\delta_s$ are the site energies, $V$ is the excitonic coupling between the two sites, $b_{s,\xi}$ and $b^\dag_{s,\xi}$ are the annihilation and creation operators for bath mode $\xi$ on site $s$, which has site-independent energy $\omega_\xi$ and linear coupling strength $g_{s,\xi}$.
We have also assumed that both donors are never simultaneously excited.
The ground state $\ket{g}$ is absent in Eq.~\eqref{eq:exciton_ham} because we take its energy to be zero and because, in the absence of a light source, it does not couple to the excited states. We set $\hbar=c=1$ throughout.

To derive a second-order Born-Markov master equation that remains valid at strong system-bath coupling, we make a unitary transformation into the polaron frame, $\tilde{H}=U_p^{\dagger}HU_p$, where $U_p=\exp\left(-\sum_{s}\ketbra{s}{s}F_s\right)$ and
\begin{equation}
    F_s=\sum_{\xi}\frac{g_{s,\xi}}{\omega_{\xi}}(b^{\dagger}_{s,\xi}-b_{s,\xi}).
\end{equation} 
The polaron transformation maps the states $\ket{L}$ and $\ket{R}$ to the same localized excitons, except with the bath modes displaced. The resulting exciton-vibrational quasiparticles are called polarons.

In the polaron frame, the Hamiltonian becomes~\cite{pollock2013multi,Lee2015}
\begin{subequations}\label{eq:polaron_ham}
\begin{align}
    \tilde{H} &= \tilde{H}^S + \tilde{H}^{SB} +\tilde{H}^{B} \\
    \tilde{H}^S &= \sum_{s\in\left\{L,R\right\}}\left(\delta_s-\Lambda_s\right)\ketbra{s}{s} \nonumber \\
    &\quad + \kappa_L\kappa_R V(\ketbra{L}{R} + \ketbra{R}{L}) \\
    \tilde{H}^{SB} &= V(\mathcal{B}_{LR}\ketbra{L}{R}+\mathcal{B}^\dag_{LR}\ketbra{R}{L}) \\
    \tilde{H}^B &= \sum_{s\in\lbrace L,R\rbrace} \sum_\xi \omega_\xi b^\dag_{s,\xi}b_{s,\xi},
\end{align}
\end{subequations}
now describing states $\ket{L}$ and $\ket{R}$ that are polarons of energy $\delta_s-\Lambda_s$ with renormalized coupling $\kappa_L\kappa_R V$.
The reorganization energy $\Lambda_s$ of the bath at site $s$ is an overall measure of the strength of system-bath interaction, and it shifts the site energies $\delta_s$ to account for the energy of the vibrational part of the polaron; it is
\begin{equation}
    \Lambda_s=\int_0^{\infty}\mathrm{d}\omega\ \frac{J_s^B(\omega)}{\omega},
\end{equation}
for a bath spectral density on site $s$ given by $J_s^B(\omega)=\sum_{\xi}|g_{s,\xi}|^2\delta(\omega-\omega_{\xi})$.
The inter-site coupling is modified by $\kappa_s \equiv \langle B^\pm_s \rangle = \exp(-\tfrac{1}{2}\phi_s(0))$, where
\begin{align}
    \phi_s(t) &= \int^\infty_0 \mathrm{d}\omega \frac{J^B_s(\omega)}{\omega^2}\left(\cos(\omega t)\coth\left(\frac{\beta\omega}{2}\right) -i\sin(\omega t)\right),\label{eq:phi}
\end{align}
and $\beta = 1/(k_B T)$.
Importantly, $0\le\kappa_s\le 1$, meaning that polaron formation always decreases inter-site coupling, the more so as $\Lambda_s$ increases.
Finally, $\mathcal{B}_{LR}=B_L^+B_R^--\kappa_L\kappa_R$ (with displacement operators $B_s^{\pm}=\exp\left(\pm F_s\right)$) describes the fluctuations of the mode displacements from their mean polaron-frame values, which induce the residual inter-site coupling scaling with $V\mathcal{B}_{LR}$ in $\tilde{H}^{SB}$.
This residual coupling is the perturbative interaction in our master equation; the fact that it scales both with the bare coupling $V$ and the bath fluctuations $\mathcal{B}_{LR}$ means that the resulting master equation is accurate so long as either $V$ or the bath fluctuations remain small. Henceforth, we assume equal exciton energies ($\delta_L=\delta_R\equiv\delta$) and system-bath couplings ($J_L^B(\omega)=J_R^B(\omega)$), which leads to $\kappa_L=\kappa_R\equiv\kappa$ and $\Lambda_L=\Lambda_R\equiv\Lambda$.

It is instructive to compare our approach with the well-known Haken-Strobl-Reineker model of coherent energy-transfer processes~\cite{haken1972coupled,haken1973exactly}. There, the bath is described classically via random white-noise fluctuations of site energies, causing pure dephasing in the site basis~\cite{wu2010efficient}. The classical noise limits this approach to infinite temperature and fails to capture the bath-induced renormalization of excitonic coupling ($V\to\kappa_L\kappa_R V$), which is essential to the results of this paper. 

Diagonalising $\tilde{H}^{S}$, we obtain the system eigenstates and eigenenergies,
\begin{align}
    \ket{\pm} &= \frac{1}{\sqrt{2}}(\ket{L}\pm\ket{R}),\\
    \delta_{\pm} &= \delta-\Lambda\pm\kappa^2V.\label{eq:eigenE}
\end{align}
Because $\delta-\Lambda$ is fixed, it is an energy offset that does not affect any results. 

The time evolution of the system's reduced density matrix, in the polaron frame and under the influence of the bath, is given by the master equation
\begin{equation}
    \dot{\rho} = -i[\tilde{H}^S,\rho] + \mathcal{L}^B\rho,\label{eq:master}
\end{equation}
where the superoperator $\mathcal{L}^B$ describes the influence of $H^{SB}$ on the system's evolution. We compute $\mathcal{L}^B$ in Supporting Information-1 (SI-1) using Redfield theory~\cite{maykuhn,breuer}, showing that, if we express the reduced density matrix in the system eigenbasis and in vectorized form $\rho= (\rho_{gg},\rho_{++},\rho_{--},\rho_{+-},\rho_{-+})^\top$, 
\begin{equation}
    \mathcal{L}^B =
    \scalemath{0.90}{
    \begin{pmatrix}
    0 & 0 & 0 & 0 & 0 \\
    0 & -k_{+-} & k_{-+} & 0 & 0 \\
    0 & k_{+-} & -k_{-+} & 0 & 0 \\
    0 & 0 & 0 & -k_d & -\frac{1}{2}(k_{+-} + k_{-+})\\
    0 & 0 & 0 & -\frac{1}{2}(k_{+-} + k_{-+})  & -k_d
    \end{pmatrix}
    }
    ,\label{eq:R_phonon}
\end{equation}
where $k_{+-}$ is the transfer rate from eigenstate $\ket{+}$ to $\ket{-}$, $k_{-+}$ is the reverse rate, $k_d = (k_{+-} + k_{-+} + 2k_p)/2$, and
$k_p$ is the pure dephasing rate of the coherences.
As shown in SI-1, these rates depend on $V$, $\kappa$, and $\phi_s(t)$.
The time-homogeneous form of Eq.~\eqref{eq:master} is a consequence of the Born-Markov approximation in Redfield theory; here, the neglect of inhomogeneous terms is particularly justified because we will only compute steady-state efficiencies, which are valid once all transients have died out.  

In the numerical results below, we always solve the full master equation~\eqref{eq:master}. However, we can obtain qualitative insight by simplifying it using the secular approximation~\cite{breuer,maykuhn}, which here means neglecting terms that couple $\rho_{+-}$ and $\rho_{-+}$ in $\mathcal{L}^B$. The approximation is valid when the two coherences are rotating rapidly in opposite directions, i.e., when $\omega_{+-} = \delta_+ - \delta_- = 2\kappa^2 V$ is much larger than other timescales of interest.
Doing so gives
\begin{equation}
\dot{\rho}_{+-} = -i\omega_{+-}\rho_{+-} -k_d\rho_{+-},
\end{equation}
which can be solved to obtain
\begin{equation}
\rho_{+-} (t)= \rho_{+-}(0)e^{-i\omega_{+-}t} e^{-k_d t}.
\label{eq:coh_func}
\end{equation}
The other coherence is $\rho_{-+}=(\rho_{+-})^*$.

We can obtain the spatial dynamics of excitons by converting from the eigenbasis to the site basis.
In particular, it can be shown that the population difference between the two sites is related to the eigenbasis coherences of Eq.~\ref{eq:coh_func}, $\rho_{LL} - \rho_{RR} = \rho_{+-} +\rho_{-+}$.
Therefore, a polaron initially localized on a single site ($\ket{L}$ for concreteness) will oscillate between the sites, with the population difference undergoing damped oscillations
\begin{equation}
    \rho_{LL}(t) - \rho_{RR}(t) = \cos(\omega_{+-} t)\exp(-k_d t),
\end{equation}
where $\omega_{+-}$ is the frequency of coherent oscillations and $k_d$ is essentially the rate of incoherent polaron diffusion.
Figure~\ref{fig2}a shows $k_d$ and $\omega_{+-}$ as functions of $\Lambda$, with $\omega_{+-}$ decreasing monotonically with $\Lambda$ and the decay rate $k_d$ increasing to a maximum before decaying.
As a result, there are three qualitatively different dynamical regimes, with example dynamics shown in Figure~\ref{fig2}b.
At low $\Lambda$, $\omega_{+-}\gg k_d$, resulting in polarons oscillating between sites before eventually relaxing into an equal mixture of site states.
At intermediate $\Lambda$, once $k_d>\omega_{+-}$, the dissipative dynamics dominates over oscillations and populations relax from a localized state to a mixture without oscillating.
Finally, large $\Lambda$ causes populations to relax more slowly, until, in the extreme $\Lambda$ limit, all population dynamics is too slow to be relevant on light-harvesting timescales.
The latter regime is important because it preserves the initial state over long times.

It may seem counterintuitive that increased coupling to the bath preserves coherent superpositions of eigenstates.
This behavior is explained by the reduction in inter-site coupling caused by polaron formation. Both $\omega_{+-}$ and $k_d$ are proportional to $\kappa$ to a power greater than one, and $\kappa$ strictly decreases with $\Lambda$ (see SI-1); since these rates cause population changes, their decrease preserves the initially localized state for longer.

\begin{table}[t]
	\centering
	\begin{tabular}{l l}
		\toprule
        Parameter & Value \\
		\midrule
		Donor energies              & $\delta_L = \delta_R = \delta = \SI{1}{eV}$     \\
		Donor excitonic coupling    & $V = \SI{5}{meV}$ \\
		Donor trans.\ dipole moment & $\mu = \SI{1.2e-28}{C.m}$ \\
		Trapping rate to acceptor   & $k_\mathrm{trap} = \SI{e12}{s^{-1}}$ \\
		Recombination rate          & $k_\mathrm{rec} = \SI{e12}{s^{-1}}$ \\
		Productive conversion rate  & $k_\mathrm{prod} = \SI{e15}{s^{-1}}$ \\
		Bath spectral density       & $J^B(\omega) = \dfrac{\Lambda}{2}\Big(\dfrac{\omega}{\omega_c}\Big)^3e^{-\omega/\omega_c}$ \\
		Bath reorganization energy  & $\Lambda = \SI{100}{meV}$ \\
		Bath cutoff frequency       & $\omega_c =\SI{20}{meV}$ \\
		Bath temperature            & $T^B =\SI{300}{K}$ \\
		Light temperature           & $T^L =\SI{6000}{K}$ \\
		\bottomrule
	\end{tabular}
	\caption{Values of parameters used in this work, except where noted otherwise.}
	\label{tab:params}
\end{table}

Having derived the effect of system-bath coupling on the donor dimer, we can now add the remaining ingredients needed to compute the effect of coherence and bath-induced noise on the efficiency of a light-harvesting system. To do so, we need a master equation that, along with the terms derived so far, includes the influences of trapping, recombination and photoexcitation.
The resulting master equation is
\begin{equation}
    \dot{\rho} = -i[\tilde{H}^S,\rho] + (\mathcal{L}^B  +\mathcal{L}^{\mathrm{rec}}+\mathcal{L}^\mathrm{trap}+\mathcal{L}^\mathrm{prod}+\mathcal{L}^{L})\rho. \label{eq:full_master}
\end{equation}

The Liouvillians $\mathcal{L}^\mathrm{rec}$ and $\mathcal{L}^\mathrm{trap}$ describe exciton loss through non-radiative recombination and trapping at the acceptor, for which we assume phenomenological forms
\begin{subequations}\label{eq:trap_recomb}
    \begin{align}
    \mathcal{L}^\mathrm{rec} \rho &= k_\mathrm{rec}\sum_{s\in\lbrace L,R\rbrace}\left(\ketbra{g}{s}\rho\ketbra{s}{g} - \tfrac{1}{2}\left\{ \ketbra{s}{s},\rho\right\}\right), \\
    \mathcal{L}^\mathrm{trap} \rho &= k_\mathrm{trap}\left(\ketbra{A}{R}\rho\ketbra{R}{A} - \tfrac{1}{2}\left\{ \ketbra{R}{R},\rho\right\}\right),
    \end{align}
\end{subequations}
where $\mathcal{L}^\mathrm{rec}$ reduces exciton population on all sites at rate $k_\mathrm{rec}$, and trapping occurs exclusively from $\ket{R}$, with exciton transfer to $\ket{A}$ at rate $k_\mathrm{trap}$.

The Liouvillian $\mathcal{L}^\mathrm{prod}$ accounts for any kind of productive or useful transfer of populations from the acceptor to the ground state, ensuring that energy can cycle through the system. It could describe return to the ground state following a charge separation event on the acceptor, or exciton transfer from the acceptor to a site outside of our system of interest.
We write it as
\begin{equation}
       \mathcal{L}^\mathrm{prod} \rho = k_\mathrm{prod}\left(\ketbra{g}{A}\rho\ketbra{A}{g} - \tfrac{1}{2}\left\{ \ketbra{A}{A},\rho\right\}\right),
        \label{eq:chargesep}
\end{equation}
where $k_\mathrm{prod}$ is the rate of acceptor-to-ground transfer.

The Liouvillians $\mathcal{L}^{\mathrm{rec}}$, $\mathcal{L}^{\mathrm{trap}}$, and $\mathcal{L}^{\mathrm{prod}}$ take the same form before and after the polaron transformation because they are simple rate processes between populations, which are invariant under the polaron transformation.

\begin{figure*}[tb]
    \centering
    \includegraphics[width=\textwidth]{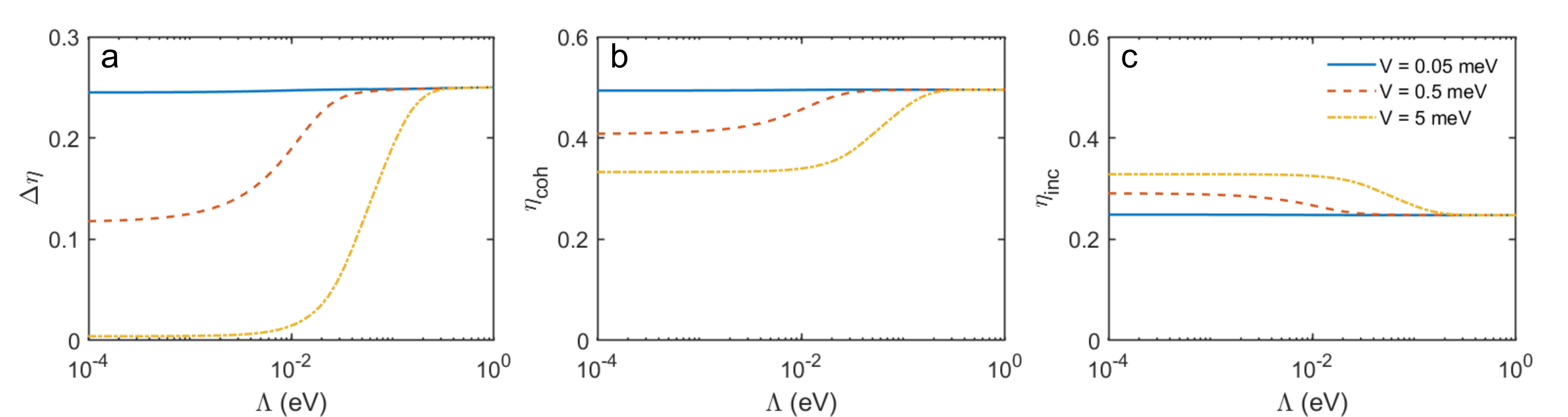}
    \caption{
    Strong system-bath coupling can amplify efficiency enhancements caused by optically induced coherence.
    \textbf{(a)}~Coherent efficiency enhancement $\Delta\eta$, computed as a function of reorganization energy $\Lambda$, is the difference $\Delta \eta = \eta_\mathrm{coh} - \eta_\mathrm{inc}$ of 
    \textbf{(b)} the efficiency $\eta_\mathrm{coh}$ when coherence is present (i.e., in polarized light) and 
    \textbf{(c)} the efficiency $\eta_\mathrm{inc}$ when coherence is absent (i.e., in unpolarized ligth).
    All three efficiencies are shown for three inter-donor couplings $V$ (legend in \textbf{(c)}).
    The increase in efficiency enhancement is caused by the suppression of inter-donor dynamics at high $\Lambda$, which increases $\eta_\mathrm{coh}$ for coherently localized excitons and decreases $\eta_\mathrm{inc}$ for excitons in an incoherent mixed state.
    Parameters as in Table~\ref{tab:params}, except for variable $\Lambda$ and $V$.
    }
    \label{fig3}
\end{figure*}

The final Liouvillian $\mathcal{L}^{L}$ describes the effect of a thermal light source on the system, derived using Hamiltonians
\begin{align}
H^L &= \sum\uql\nu\uq\label{eq:Hpt} a^{\dagger}\uql a\uql,\\
H^{SL} &= -\sum_{s\in\{L,R\}}\mb{d}_s\cdot\mb{E},\label{eq:HptSO}
\end{align}  
where $H^L$ is the energy of the photonic degrees of freedom and $H^{SL}$ the interaction between the system and the light. The photons have wavenumber $\mb{q}$, polarization $\lambda$ and energy $\nu\uq$, described by the harmonic operators $a\uql$. The transition dipole operator of site $s$ is $\mb{d}_s=\bm{\mu}_s\left(\ketbra{s}{g}+\ketbra{g}{s}\right)$, where $\bm{\mu}_s$ is the transition dipole moment. The system-light interaction is between the dipole operators and the electric field, which, in the electric dipole approximation, is
\begin{equation}
\mb{E}=-i\sum\uql f\uq\bm{\epsilon}\uql(a^{\dagger}\uql-a\uql),
\end{equation}
where $f\uq=\sqrt{\nu\uq/(2\mathcal{V})}$ is the electric-field amplitude of mode $\mb{q}$ with quantization volume $\mathcal{V}$ and $\bm{\epsilon}\uql$ for $\lambda=1,2$ are the two polarization modes perpendicular to $\mb{q}$. 

We assume that the light is perpendicular to the plane containing both transition dipole vectors, and that these couple to the field with equal strength. Therefore, we set $\bm{\mu}_L=\mu\mb{e}_x$, $\bm{\mu}_R=\mu\mb{e}_y$ and  $\mb{q}=q\mb{e}_z$. When the light is polarized parallel to the right dipole, $\bm{\epsilon}_{\mb{q}1}=\mb{e}_y$ and $\bm{\epsilon}_{\mb{q}2}=0$. In unpolarized light, the field can oscillate in any direction perpendicular to $\mb{q}$, meaning that we can write $\bm{\epsilon}_{\mb{q}1}=\cos\varphi\uq\mb{e}_x+\sin\varphi\uq\mb{e}_y$ and $\bm{\epsilon}_{\mb{q}2}=-\sin\varphi\uq\mb{e}_x+\cos\varphi\uq\mb{e}_y$, where $\varphi\uq$ is the azimuthal angle in the spherical polar coordinate system in which $\mb{q}$ defines the z-axis. The transition rates then depend on values averaged over all $\varphi\uq$.

To account for the strong system-bath coupling in the dimer, we must apply the polaron transformation to the dipole operators in $H^{SL}$, replacing them with
\begin{equation}
\tilde{\mb{d}}_s = U^{\dagger}_p\mb{d}_s U_p = \bm{\mu}_s(B_s^-\ketbra{g}{s}+B_s^+\ketbra{s}{g}). \label{eq:dipolepolaron}
\end{equation}
The operator $B_s^+\ketbra{s}{g}$ causes simultaneous creation of an exciton and displacement of the bath modes and so describes the creation of a polaron. Likewise, $B_s^-\ketbra{g}{s}$ destroys a polaron. 

Calculating $\mathcal{L}^{L}$ is more difficult in the polaron frame than for a bare system interacting with light. In SI-2, we show that, within the Born-Markov approximation and for thermal light,
\begin{equation}\label{eq:Lpt}
\mathcal{L}^{L}=
\scalemath{0.9}{
\begin{pmatrix}
-(\gamma_+^{\uparrow}+\gamma_-^{\uparrow}) & \gamma_+^{\downarrow} & \gamma_-^{\downarrow} & \theta^{\downarrow}_++\theta^{\downarrow}_- & \theta^{\downarrow}_++\theta^{\downarrow}_-\\
\gamma_+^{\uparrow} & -\gamma_+^{\downarrow} & 0 & -\theta^{\downarrow}_- & -\theta^{\downarrow}_-\\
\gamma_-^{\uparrow} & 0 & -\gamma_-^{\downarrow} & -\theta^{\downarrow}_+ & -\theta^{\downarrow}_+\\
\theta^{\uparrow}_++\theta^{\uparrow}_- & -\theta^{\downarrow}_+ & -\theta^{\downarrow}_- & -\frac{1}{2}(\gamma_+^{\downarrow}+\gamma_-^{\downarrow}) & 0\\
\theta^{\uparrow}_++\theta^{\uparrow}_- & -\theta^{\downarrow}_+ & -\theta^{\downarrow}_- & 0 & -\frac{1}{2}(\gamma_+^{\downarrow}+\gamma_-^{\downarrow})
\end{pmatrix}
},
\end{equation}
where the basis is the same as in Eq~\eqref{eq:R_phonon}. The rates for the optical excitation and radiative recombination of the polaron states are, respectively, $\gamma_{\pm}^{\uparrow}=\gamma(-\delta_{\pm})$ and $ \gamma_{\pm}^{\downarrow}=\gamma(\delta_{\pm})$,
where $\delta_{\pm}$ are defined in Eq.~\eqref{eq:eigenE} and
\begin{equation}
    \gamma(\omega)=\pi(G_{RR}+G_{LL})\mathcal{F}(\omega).\label{eq:optical}
\end{equation}
The function $\mathcal{F}(\omega)$, given in SI-2, depends on the system's coupling to both the light and the bath, and
\begin{equation}\label{eq:polFactor}
G_{ss'}=2\int_0^{2\pi}\mathrm{d}\varphi\uq \sum_{\lambda}\left(\bm{\mu}_{s}\cdot\bm{\epsilon}\uql\right)\left(\bm{\mu}_{s'}\cdot\bm{\epsilon}\uql\right),
\end{equation} 
captures the interplay of the polarization with the orientation of the dipoles. For polarized light, $G_{RR}=4\pi\mu^2$ and $G_{LL}=0$, whilst for unpolarized light, $G_{RR}=G_{LL}=4\pi\mu^2$. The light also pumps and decoheres coherences between the polaron states at rates $\theta_{\pm}^{\uparrow}=\theta(-\delta_{\pm})$ and $ \theta_{\pm}^{\downarrow}=\theta(\delta_{\pm})$, respectively, where
\begin{equation}
    \theta(\omega)=\frac{\pi}{2}\left(G_{LL}-G_{RR}\right)\mathcal{F}(\omega).\label{eq:cohgen}
\end{equation}

Eqs.~\eqref{eq:polFactor} and ~\eqref{eq:cohgen} reveal the crucial difference between polarized and unpolarized light: because the coherence pumping rate $\theta(\omega)$ depends on $G_{LL}-G_{RR}$, coherences between the eigenstates are only pumped by the polarized light, while unpolarized light only pumps populations.

Figure~\ref{fig2}c shows the eigenstate coherence as a function of time for the optically driven system initialized in the ground state. In all three cases, the light is assumed polarized to ensure that the coherence is pumped. As predicted, stronger system-bath coupling ensures that the coherences are both larger and longer lived. This result is caused by the behavior of both the optical pumping and dephasing as functions of system-bath coupling strength. The coherence pumping rate $\theta(\delta_\pm)$ increases with the system-bath coupling because of its dependence on $\mathcal{F}(\omega)$, which is an increasing function of system-bath coupling for the parameters considered in this paper (see SI-2). By contrast, dephasing occurs at rate $(k_{+-} + k_{-+})/2$, which, scaling as $(\kappa^2 V)^2$, is weaker for stronger system-bath coupling (see SI-1). The overall effect is that, at higher system-bath coupling, the light is able to pump coherences faster and they decay slower, allowing them to grow in size and persist for longer.

In our polaron framework, the optical generation of polarons and coherences between them is assumed to be instantaneous. More realistically, light would create an exciton, which would then relax into a polaron, a process that can create bath-induced coherences~\cite{ma2015forster,ma2015forster2}. However, the two processes usually happen on very different timescales, with the latter being much faster; as a result, they can be considered as one for many purposes, without the need to dynamically resolve the polaron formation.

We now use the full master equation in Eq.~\eqref{eq:full_master} to simulate the full, steady-state, light-harvesting process under excitation by both the polarized and unpolarized light sources.
The figure of merit in our comparisons is the efficiency, the proportion of photogenerated polarons that reach the acceptor instead of recombining. At steady state, we calculate the efficiency as the ratio between the population flux from the acceptor to the ground state and the flux out of the ground state due to optical pumping,
\begin{equation}
    \eta = \frac{k_\mathrm{prod}\rho_{AA}^\mathrm{ss}}{(\gamma_+^{\uparrow}+\gamma_-^{\uparrow})\rho_{gg}^\mathrm{ss}},
\end{equation}
where $\rho^\mathrm{ss}$ is the steady-state density matrix. We can obtain $\rho^\mathrm{ss}$ in two ways. The first approach is to initialize the system in the ground state, and then evolve it under Eq.~\eqref{eq:full_master} until the density matrix converges. The second approach is to find $\rho^\mathrm{ss}$ as the unique element of the nullspace of the Liouvillian operator that occurs on the right-hand side of Eq.~\eqref{eq:full_master}.
Another figure of merit is $\Delta\eta = \eta_{\mathrm{coh}} - \eta_{\mathrm{inc}}$, the difference between the efficiencies under polarized ($\eta_{\mathrm{coh}}$) and unpolarized ($\eta_{\mathrm{inc}}$) excitation, to indicate how much coherence enhances the efficiency.

\begin{figure*}[tb]
    \centering
    \includegraphics[width=0.8\textwidth]{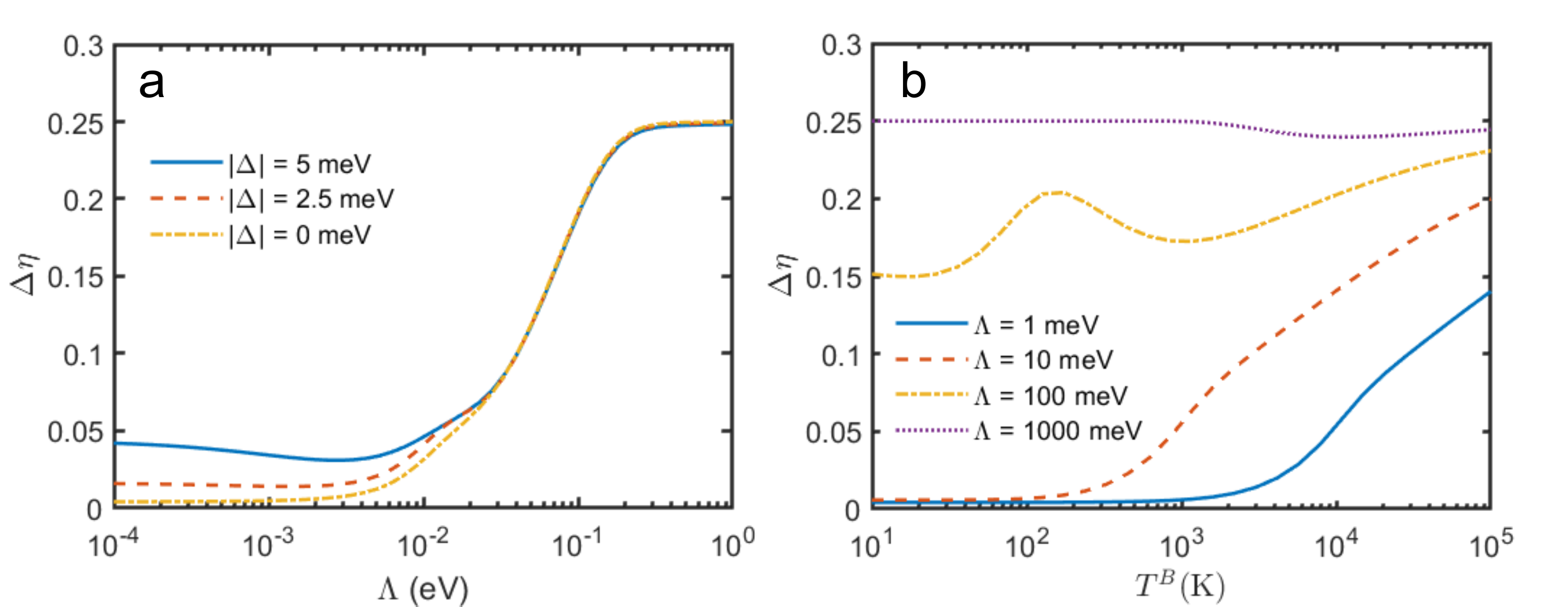}
    \caption{ 
    \textbf{(a)} Efficiency enhancement $\Delta\eta$ as a function of $\Lambda$, for several detunings $\Delta$ between donor sites (the plots are independent of the sign of $\Delta$).
    Detuning can increase $\Delta\eta$ at small $\Lambda$, but has a negligible effect at large $\Lambda$.
    \textbf{(b)} $\Delta\eta$ as a function of bath temperature $T^B$, for several values of $\Lambda$. The complicated behavior, discussed in the text, indicates that increasing $T^B$ is not always comparable to increasing $\Lambda$; while larger $\Lambda$ always increases $\Delta\eta$, the same is not true for larger $T^B$.
    Parameters as in Table~\ref{tab:params}, except for variable $\Lambda$, $\Delta$, and $T^B$.}
    \label{fig4}
\end{figure*}

For numerical calculations, we used the values shown in Table~\ref{tab:params}, which are inspired by biological light-harvesting complexes.
In particular, the choice of $\omega_c$ is close to estimates for photosynthetic systems~\cite{pollock2013multi,Pachon2017}, and its being significantly larger than $V$ ensures the accuracy of the polaron-transformed model~\cite{mccutcheon2011general,pollock2013multi}. 
$k_\mathrm{trap}$ and $k_\mathrm{rec}$ were chosen to be equal to reduce the number of free parameters, and their value corresponds to typical exciton transfer rates in biological systems~\cite{Leon-Montiel2014,Tscherbul2018}.
$k_\mathrm{prod}$ is much larger than all other rates to keep the ground state $\rho_{gg}$ mostly undepleted throughout the process; indeed, $k_\mathrm{prod}$ does not affect the efficiency, as it only serves to transfer excitons from the acceptor to ground, after trapping has already occurred.
Finally, $T^B$ and $T^L$ reflect, respectively, room temperature and the temperature of solar radiation.

Figure~\ref{fig3}a shows our main result, that $\Delta\eta$ always increases with $\Lambda$---i.e., that stronger coupling to the bath further improves efficiency enhancements due to coherence---and that the effect is more pronounced at larger $V$. The improvement is a consequence of the fact that stronger system-bath coupling inhibits dynamics within the donor dimer, as discussed above and shown in Figure~\ref{fig2}a. However, this inhibition exerts opposite effects on the efficiencies under polarized and unpolarized excitation, as shown in Figures~\ref{fig3}b and \ref{fig3}c. In polarized light, increasing $\Lambda$ increases the efficiency $\eta_{\mathrm{coh}}$ because the suppression of dynamics ensures that the exciton remains on the initial site $\ket{R}$, from where it can be trapped by the acceptor. In the limit of infinite $\Lambda$, $\ket{L}$ does not participate in the process at all, and the efficiency is the branching ratio of polarons being trapped from $\ket{R}$ as opposed to decaying to ground; in our computations, $k_R=k_\mathrm{trap}$ implies that the limiting $\eta_{\mathrm{coh}}$ is 0.5.
By contrast, in unpolarized light, which pumps an incoherent mixture of the two site states with equal populations, increasing $\Lambda$ decreases the efficiency $\eta_{\mathrm{inc}}$. In that case, the suppression of dynamics at high $\Lambda$ prevents the exchange of populations between sites, meaning that the fraction of polarons initially on $\ket{L}$ cannot be trapped by the acceptor and will be lost to recombination.
With half of the polarons beginning on $\ket{R}$ and half of those lost to recombination (as in the polarized case), the efficiency at infinite $\Lambda$ is 0.25.

The maximum possible enhancement (no longer assuming $k_\mathrm{trap}=k_\mathrm{rec}$) is $\Delta\eta=0.5$, just as in the weak-coupling case~\cite{Tomasi2019}. It occurs in the limit of infinite $k_\mathrm{trap}/k_\mathrm{rec}$ and $\Lambda$, where all population on $\ket{R}$ is trapped, giving $\eta_{\mathrm{coh}}=1$, twice as much as $\eta_{\mathrm{inc}}=0.5$.

Although we have assumed degenerate donor sites so far, our conclusions are robust to small detunings that might be caused by experimental imperfections. Figure~\ref{fig4}a shows how the behavior of $\Delta\eta$ as a function of $\Lambda$ is affected by a range of detunings $\Delta = \delta_R - \delta_L$. These results were obtained using the full expressions in SI-1 and SI-2, as opposed to the simplified degenerate versions given here. Because the site transition dipole moments remain perpendicular, the light excites the same states as in the degenerate case (i.e., localized on $R$ for polarized light, or an equal mixture of site states for unpolarized light).
In particular, Figure~\ref{fig4}a shows that $\Delta$ does not markedly affect results at high $\Lambda$, because the motion of initial states remains suppressed, giving the same limiting efficiencies as in the degenerate model. At low $\Lambda$, the slightly asymmetric dynamics induced by $\Delta$ means that, in the limit of rapid exciton motion, the populations on $\ket{L}$ and $\ket{R}$ will be unequal, yielding a small, non-zero value of $\Delta\eta$.
Overall, the most important result of Figure~\ref{fig4}a is that detuning is not detrimental to demonstrations of coherence-enhanced light-harvesting, meaning that an experimental demonstration does not require perfect degeneracy.

In Figure~\ref{fig4}a, we introduced detunings with a maximum magnitude of $|\Delta| = V$, where the eigenstates $\ket{\pm}$ are $72:28$ superpositions of the two site states. A substantially larger detuning might lead to similar results, but it would become difficult to assign efficiency enhancements to eigenbasis coherence. In the limit of high detuning, the eigenbasis coincides with the site basis, meaning that the localized states prepared by polarized light would lack eigenbasis coherences too.

Throughout this paper, the reorganization energy $\Lambda$ has set the overall strength of system-bath coupling. Sometimes, the temperature $T^B$ has a similar effect on open-system dynamics as $\Lambda$, because both can be thought of as increasing the perturbations on the system. However, this is a rough analogy, and, while being straightforward to do experimentally, increasing $T^B$ is not a reliable way to achieve the same effects as increasing $\Lambda$.
In our system, the two parameters have similar effects on $\Delta \eta$ only in some regimes. As shown in Figure~\ref{fig4}b, increasing $T^B$ increases $\Delta \eta$, but only for small $\Lambda$; at larger $\Lambda$, the temperature dependence of $\Delta \eta$ is complicated. We cannot offer simple explanations of all the features because $T^B$ affects the dynamics in multiple, often-competing ways that can operate on similar energy scales. For example, while $\phi_s(t)$ is simply proportional to $\Lambda$, it does not have a straightforward dependence on $T^B$; furthermore, the calculations in SI-1 show that transfer rates $k_{\pm\mp}$ depend on complicated functions of $\phi_s(t)$ in a way that also depends on the energies $\delta_\pm$, which themselves are modified by $T^B$. Therefore, while in many cases a coherent enhancement of light-harvesting efficiency could be increased by heating up the system, this outcome is not guaranteed.

Our results show that coherent enhancements of light harvesting can not only be robust to noise, but that system-bath coupling can increase them by suppressing polaron dynamics, resulting in significantly improved efficiency over broad parameter ranges. 
Unlike in most quantum technologies where system-bath coupling is deleterious, our system could be realized in any naturally noisy material where three sites---whether molecules, artificial atoms, or semiconductor nanostructures---can be coupled in the geometry of Figure~\ref{fig1}. 
Thus, the benefit of site-basis noise significantly broadens the range of systems that can be considered for demonstrating coherent light-harvesting enhancements.

Although our scheme is a fundamental, proof-of-principle proposal for conclusively demonstrating advantages of coherence, it also opens up the promise of practical applications.
In the future, similar arguments or other ways for controlling coherence could be extended to improve the performance of larger systems, including those with more complicated spatial and energetic arrangements of chromophores or with more complex ways of taking advantage of engineered coherent effects.

\section*{Acknowledgements} 
S.T. and I.K. were supported by a Westpac Scholars Trust Research Fellowship, by an Australian Government Research Training Program scholarship, by the University of Sydney Nano Institute Grand Challenge \textit{Computational Materials Discovery}, and by computational resources from the National Computational Infrastructure and the Sydney Informatics Hub. D.M.R. acknowledges studentship funding from EPSRC grant EP/L015110/1, and E.M.G. the EPSRC grant EP/T007214/1.

\section*{Supporting Information} 
Supporting information is available on the derivations of the non-secular bath and light Liouvillians.


\providecommand{\latin}[1]{#1}
\makeatletter
\providecommand{\doi}
  {\begingroup\let\do\@makeother\dospecials
  \catcode`\{=1 \catcode`\}=2 \doi@aux}
\providecommand{\doi@aux}[1]{\endgroup\texttt{#1}}
\makeatother
\providecommand*\mcitethebibliography{\thebibliography}
\csname @ifundefined\endcsname{endmcitethebibliography}
  {\let\endmcitethebibliography\endthebibliography}{}

\end{document}


\title{Supporting Information: Environmentally Improved Coherent Light Harvesting}

\author{Stefano Tomasi}
\affiliation{School of Chemistry and University of Sydney Nano Institute, University of Sydney NSW 2006, Australia}

\author{Dominic M. Rouse}
\affiliation{SUPA, School of Physics and Astronomy, University of St Andrews, St Andrews KY16 9SS, UK}

\author{Erik M. Gauger}
\affiliation{SUPA, Institute of Photonics and Quantum Sciences, Heriot-Watt University, Edinburgh EH14 4AS, UK}

\author{Brendon W. Lovett}
\affiliation{SUPA, School of Physics and Astronomy, University of St Andrews, St Andrews KY16 9SS, UK}

\author{Ivan Kassal}
\email[Email: ]{ivan.kassal@sydney.edu.au}
\affiliation{School of Chemistry and University of Sydney Nano Institute, University of Sydney NSW 2006, Australia}

\maketitle

\section*{SI-1. Bath Liouvillian}

Here, we derive Eq.~(9) of the main text using Redfield theory, largely following refs.~\cite{Lee2015,nazir2016modelling,pollock2013multi}.
In general, for a weak system-environment interaction Hamiltonian of the form $H^{I} = \sum_u K^{(u)}\Phi^{(u)}$, with system operators $K^{(u)}$ and environment operators $\Phi^{(u)}$, Redfield (or second-order Born-Markov) theory predicts that the reduced density matrix of the system evolves as~\cite{maykuhn,breuer}
\begin{equation}
    \dot{\rho} = -i[H^\mathrm{sys},\rho] + \mathcal{L}^I\rho,
\end{equation}
where the superoperator $\mathcal{L}^I$ acts on $\rho$ as follows:
\begin{equation}
\mathcal{L}^I\rho = \sum\limits_{a,b,c,d} \Gamma^I_{ab,cd}(\omega_{dc})\big(A^{(cd)}\rho A^{(ab)} - A^{(ab)}A^{(cd)}\rho\big)
+ \Gamma^I_{dc,ba}(\omega_{ab})\big(A^{(cd)}\rho A^{(ab)} - \rho A^{(ab)}A^{(cd)}\big),\label{eq:redf}
\end{equation}
where $A^{(ab)} = \ketbra{a}{b}$ and
\begin{equation}
\Gamma_{ab,cd}^I(\omega)= \sum_{u,v} K^{(u)}_{ab}K^{(v)}_{cd}\Re \int^\infty_0 \mathrm{d}t \langle \bar{\Phi}^{(u)}(t)\bar{\Phi}^{(v)}(0)\rangle e^{i\omega t}, \label{eq:gamma}
\end{equation}
with indices $a,b,c,d,e$ representing eigenstates of the system Hamiltonian $H^\mathrm{sys}$ and the bar indicating an operator in the interaction picture. 
We have taken the real part of Eq.~\eqref{eq:gamma} to ignore Lamb shifts, which are usually small and can in any case be incorporated into the system Hamiltonian~\cite{breuer}.
In the eigenbasis, $K^{(u)}=\sum_{a,b}K_{ab}^{(u)}\ketbra{a}{b}$.

In our case, $H^\mathrm{sys}$ is the polaron-frame Hamiltonian $\tilde{H}^S$ and $H^{I}$ is the polaron-frame system-bath interaction
\begin{equation}
    \tilde{H}^{SB} = V(\mathcal{B}_{LR}\ketbra{L}{R}+\mathcal{B}^\dag_{LR}\ketbra{R}{L}),
\end{equation}
which allows us to set $K^{(st)} = \ketbra{s}{t}$ and $\Phi^{(st)} = \mathcal{B}_{st}$, where $st,uv\in\lbrace LR,RL\rbrace$.
Eq.~\eqref{eq:gamma} then becomes
\begin{equation}
    \Gamma^B_{ab,cd}(\omega) = V^2 \sum_{s,t,u,v} c^{(s)}_a c^{(t)*}_b c^{(u)}_c c^{(v)*}_d  M^B_{stuv}(\omega),
\end{equation}
where the label $B$ indicates the bath, $c^{(s)}_a = \braket{a|s}$, and 
\begin{equation}
     M^B_{stuv}(\omega)=\Re\int^\infty_0 \mathrm{d}t \langle \bar{\mathcal{B}}_{st}(t)\bar{\mathcal{B}}_{uv}(0)\rangle e^{i\omega t}.
\end{equation}
It can be shown that $\langle \bar{\mathcal{B}}_{st}(t)\bar{\mathcal{B}}_{uv}(0)\rangle = \kappa^4(e^{-\lambda_{stuv}\phi(t)} - 1)$,
where $\lambda_{stuv} = \delta_{su} +\delta_{tv}-\delta_{sv}-\delta_{tu}$~\cite{Lee2015,nazir2016modelling}.

Redfield theory is expressed in the system eigenbasis, in this case
\begin{equation}
\begin{pmatrix}
\ket{+}\\\ket{-}
\end{pmatrix}
=
\begin{pmatrix}
\cos(\chi/2) & \sin(\chi/2) \\
-\sin(\chi/2) & \cos(\chi/2)
\end{pmatrix}
\begin{pmatrix}
\ket{R}\\\ket{L}
\end{pmatrix},
\end{equation}
where $\cos\chi=\Delta'/\Omega$, $\sin\chi=V'/\Omega$, $\Omega=\sqrt{\Delta'^2+V'^2}$, $V'=\kappa_L\kappa_R V$, $\Delta'=\delta_R'-\delta_L'$ and $\delta_s'=\delta_s-\Lambda_s$ and $\Lambda_s$ is the reorganization energy at site $s$.

If we represent the reduced density matrix in vector form, $\rho= (\rho_{++},\rho_{--},\rho_{+-},\rho_{-+})^\top$, $\mathcal{L}^B$ becomes
\begin{equation}
    \mathcal{L}^B =
    \begin{pmatrix}
    -k_{+-} & k_{-+} & \sigma(0)& \sigma(0) \\
    k_{+-} & -k_{-+} & -\sigma(0) & -\sigma(0) \\
    \sigma(\omega_{+-}) & \sigma(\omega_{-+}) & -k_d & Z \\
    \sigma(\omega_{+-}) & \sigma(\omega_{-+}) & Z & -k_d
    \end{pmatrix},\label{eq:R_phonon2}
\end{equation}
with the individual rates
\begin{align}
    k_{+-} &= 2V^2\big[(p_1^2+p_2^2) M^B_{LRRL}(\omega_{+-})-2 p_1 p_2 M^B_{LRLR}(\omega_{+-})\big],\\
    k_{-+} &= \exp\left(-\beta\omega_{+-}\right)k_{+-},\\
    k_d &= \tfrac{1}{2}(k_{+-} +k_{-+} +2k_p), \\
    k_p &= 8V^2p_1p_2\left[ M^B_{LRLR}(0)+ M^B_{LRRL}(0)\right],\\
    \sigma(\omega)&= 2V^2\sqrt{p_1p_2}(p_2-p_1)( M^B_{LRLR}(\omega)+ M^B_{LRRL}(\omega)), \\
    Z &= V^2\big[(p_1^2+p_2^2)( M^B_{LRLR}(\omega_{+-})+ M^B_{LRLR}(\omega_{-+}))-2 p_1 p_2( M^B_{LRRL}(\omega_{+-})+ M^B_{LRRL}(\omega_{-+}))\big],
\end{align}
where $p_1 = \sin^2(\chi/2)$ and $p_2 = \cos^2(\chi/2)$.

For degenerate donors ($\delta_L=\delta_R$), $p_1 = p_2 = 1/2$ and the equations simplify to
\begin{align}
    k_{+-} &= 2\kappa^4 V^2\Re\int_0^\infty \mathrm{d}t \sinh(2\phi(t))e^{i\omega_{+-}t}, \\
    k_p &= 4\kappa^4 V^2\Re\int_0^\infty \mathrm{d}t (\cosh(2\phi(t))-1), \\
    \sigma(\omega) &= 0,\label{eq:sigma_zero} \\
    Z &= -\tfrac{1}{2}(k_{+-}+k_{-+}),
\end{align}
yielding Eq.~(9) of the main text. In particular, the vanishing of $\sigma(\omega)$ implies that, in this special case, the evolution of the coherences is decoupled from that of the populations.

\section*{SI-2: Light Liouvillian}

Here, we derive the light Liouvillian $\mathcal{L}^L$ using Redfield theory. For completeness, our derivation is general, and we will simplify it to the form given in Eq.~(20) of the main text at the end, by assuming the electric dipole approximation and degenerate donor sites with perpendicular transition dipole moments and identical baths.

We follow the same Redfield treatment as in SI-1, except that the interaction Hamiltonian is now the system-light Hamiltonian in the polaron frame. 
From Eqs.~(17) and~(19) of the main text, it can be written as
\begin{equation}
	\tilde{H}^{SL}=-\sum_{s\in\{L,R\}}\sum\uql (B^+_s\ketbra{s}{g}+B_s^-\ketbra{g}{s})f\uq \mu_{s\mb{q}\lambda} E_{s\mb{q}\lambda},
\end{equation}
where $E_{s\mb{q}\lambda}=-i(a^{\dagger}\uql e^{-i\mb{q}\cdot\mb{r}_s}-a\uql e^{i\mb{q}\cdot\mb{r}_s})$,  
$\mu_{s\mb{q}\lambda}=\bm{\mu}_{s}\cdot\bm{\epsilon}\uql$, and $\mb{r}_s$ is the spatial position of site $s$.

As in SI-1, we can factorize $\tilde{H}^{SL}$ into system operators $K^{(u)}$ and environment operators $\Phi^{(u)}$.
There are four system operators, $K^{(sg)}=\ketbra{s}{g}$ and $K^{(gs)}=\ketbra{g}{s}$ for $s\in\{L,R\}$, and four corresponding environment operators:
\begin{align}
    \Phi^{(sg)} &= -\sum\uql f\uq \mu_{s\mb{q}\lambda} E_{s\mb{q}\lambda} B_s^+,\\
    \Phi^{(gs)} &= -\sum\uql f\uq \mu_{s\mb{q}\lambda} E_{s\mb{q}\lambda} B_s^-,
\end{align}
which contain both bath and light operators. The master equation Eq.~\eqref{eq:redf} now gives the evolution due to the light,
\begin{equation}
\mathcal{L}^L\rho = \sum\limits_{\substack{a,b\;\in\\\{+,-\}}} \Gamma^L_{ga,bg}(\omega_{gb})\big(A^{(bg)}\rho A^{(ga)} - A^{(ga)}A^{(bg)}\rho\big)
+ \Gamma^L_{ag,gb}(\omega_{bg})\big(A^{(gb)}\rho A^{(ag)} - A^{(ag)}A^{(gb)}\rho\big)+\mathrm{h.c.} ,\label{eq:redfL}
\end{equation}
where h.c. denotes the Hermitian conjugate and 
\begin{equation}
    \Gamma_{ab,cd}^L(\omega)=
    \sum\limits_{\substack{u,v\;\in\\\{Lg,gL,Rg,gR\}}} K^{(u)}_{ab}K^{(v)}_{cd}\mathrm{Re}\int^\infty_0 \mathrm{d}t \langle \bar{\Phi}^{(u)}(t)\bar{\Phi}^{(v)}(0)\rangle e^{i\omega t}. \label{eq:gammaL}
\end{equation}
 
In Eq.~\eqref{eq:redfL}, we omitted any terms that only contribute to the evolution of $\rho_{g\pm}$ and $\rho_{\pm g}$ (those proportional to $\Gamma^L_{ga,gb}$ and $\Gamma^L_{ag,bg}$) because these are decoupled from the elements we are interested in, namely $\rho_{gg}$, $\rho_{++}$, $\rho_{--}$, $\rho_{+-}$ and $\rho_{-+}$. 
Using the expressions for $K_{ab}^{(u)}$ and $\Phi^{(u)}$, the transition rates in Eq.~\eqref{eq:redfL} become
\begin{subequations}\label{eq:gammaL2}
\begin{align}
    \Gamma_{ga,bg}^L(\omega)=\sum_{s,s'\in\{L,R\}}c_a^{(s)*}c_b^{(s')}M_{ss';+-}^{L}(\omega),\\
    \Gamma_{ag,gb}^L(\omega)=\sum_{s,s'\in\{L,R\}}c_a^{(s)}c_b^{(s')*}M_{ss';-+}^{L}(\omega),
\end{align}
\end{subequations}
where $c_a^{(s)}=\braket{a|s}$ and, within the Born approximation,
\begin{equation}\label{eq:Mssp}
 M_{ss';\pm\mp}^{L}(\omega)=\mathrm{Re}\int_0^{\infty}\mathrm{d}t\ e^{i\omega t}\sum\uql\sum_{\mb{q}'\lambda'}f\uq f_{\mb{q}'}\mu_{s\mb{q}\lambda}\mu_{s'\mb{q}'\lambda'}\langle \bar{E}_{s\mb{q}\lambda}(t) \bar{E}_{s'\mb{q}'\lambda'}(0)\rangle_{L}\langle \bar{B}^\pm_s(t) \bar{B}^\mp_{s'}(0)\rangle_{B}.
\end{equation}
The Born approximation assumes the light and bath are uncorrelated, i.e., that the environment density matrix can be factorized as $\rho^{B,L}=\rho^B\rho^L$, allowing the factorization of the environment expectation values in Eq.~\eqref{eq:Mssp}. 

The bath expectation values are~\cite{nazir2016modelling}
\begin{equation}
\langle \bar{B}_s^{\pm}(t) \bar{B}_{s'}^{\mp}(0)\rangle_B=\begin{cases}
\kappa_s\kappa_{s'} &\text{for }s\neq s'\\
\kappa_s^2e^{\phi_s(t)} &\text{for }s=s',
\end{cases}\label{eq:BB}
\end{equation}
where $\phi_s(t)$ is given in Eq.~(5) of the main text. Because $\langle \bar{B}_s^{-}(t) \bar{B}_{s'}^{+}(0)\rangle_B=\langle \bar{B}_s^{+}(t) \bar{B}_{s'}^{-}(0)\rangle_B$ and $c_a^{(s)}$ are real valued, it follows that in Eqs.~\eqref{eq:gammaL2} $M^L_{ss';+-}(\omega)=M^L_{ss';-+}(\omega)$ and therefore  $\Gamma^L_{ga,bg}(\omega)=\Gamma^L_{ag,gb}(\omega)$. Hence, we will denote these as $M_{ss'}^L(\omega)$ and $\Gamma^{L}_{ab}(\omega)$, respectively, related by
\begin{equation}
    \Gamma^L_{ab}(\omega)=\sum_{s,s'\in\{L,R\}}c_a^{(s)}c_b^{(s')}M_{ss'}^L(\omega).
\end{equation}

The light expectation values are found using the standard bosonic expectation values for thermal light, \begin{align}
    &\langle a^{\dagger}\uql a_{\mathbf{q}'\lambda'} \rangle_{L}=N^L(\nu\uq)\delta_{\mb{q}\mb{q}'}\delta_{\lambda\lambda'},\\
    &\langle a\uql a_{\mathbf{q}'\lambda'}^{\dagger} \rangle_{L}=(1+N^L(\nu\uq))\delta_{\mb{q}\mb{q}'}\delta_{\lambda\lambda'},
\end{align}
and otherwise zero where $N^L(\nu)$ is the Bose-Einstein distribution of the light. Then, we use one of the $\mathbf{q}\lambda$ sums in Eq.~\eqref{eq:Mssp} to evaluate the $\delta$-functions and take the continuum limit of the other sum by setting $\nu\uq\to\nu$ and
\begin{equation}
    \sum\uq \to\frac{\mathcal{V}}{(2\pi)^3}\int_0^{\infty}\mathrm{d}\nu\ \nu^2\int_{\Omega\uq}\mathrm{d}\Omega\uq, 
\end{equation}
where $\Omega\uq=(\theta\uq,\phi\uq)$ is the solid angle over which the field modes are distributed~\cite{ficek2005quantum}. The result is
\begin{equation}\label{eq:Iss}
M^L_{ss'}(\omega)=\int_0^{\infty}\mathrm{d}\nu\ J^L(\nu)\left[G_{ss'}(-\nu)N^L(\nu)\mathcal{K}_{ss'}(\omega+\nu)
+G_{ss'}(\nu)(1+N^L(\nu))\mathcal{K}_{ss'}(\omega-\nu)\right],
\end{equation}
where the spectral density of the light is $J^L(\nu)=\nu^3/(16\pi^3)$. The bath influence is contained within
\begin{equation}\label{eq:vibInt}
    \mathcal{K}_{ss'}(\omega)=\mathrm{Re}\int_0^{\infty}\mathrm{d}t\ \langle \bar{B}_s^+(t) \bar{B}^-_{s'}(0)\rangle_{B}e^{i\omega t},
\end{equation}
and the polarization factors are
\begin{equation}\label{eq:polFactorSI}
    G_{ss'}(\nu)=\int_0^{\pi}\mathrm{d}\theta\uq\ \sin\theta\uq\int_0^{2\pi}\mathrm{d}\varphi\uq\sum_{\lambda}\left(\bm{\mu}_{s}\cdot\bm{\epsilon}\uql\right)\left(\bm{\mu}_{s'}\cdot\bm{\epsilon}\uql\right)e^{ i\nu\mb{e}_{\mb{q}}\cdot\left(\mb{r}_s-\mb{r}_{s'}\right)},
\end{equation}
where we have written $\mb{q}=\left|\mb{q}\right|\mb{e}\uq=\nu\mb{e}\uq$.

The polarization factors $G_{ss'}(\nu)$ can be simplified for the dimer geometry in the main text, where $\mb{r}_s-\mb{r}_{s'}$ is perpendicular to $\bm{\epsilon}\uql$, making the exponential in Eq.~\eqref{eq:polFactorSI} equal to 1. This makes the polarization factors independent of frequency (as in the main text in Eq.~(22)) even if the electric dipole approximation is not made.

The bath influence on the optical rates is contained within $\mathcal{K}_{ss'}(\omega)$, which can be evaluated for $s\neq s'$ using the identity
\begin{equation}
    \mathrm{Re}\int_0^{\infty}\mathrm{d}t\ e^{i\omega t}=\pi\delta(\omega),
\end{equation}
and Eq.~\eqref{eq:BB} to give
\begin{equation}
    \mathcal{K}_{ss'}(\omega)\big\vert_{s\neq s'}=\pi \kappa_s\kappa_{s'}\delta(\omega).
\end{equation}
The $s=s'$ case is more complicated, but an analytic expression for $\mathcal{K}_{ss'}(\omega)$ can be found by assuming that the bath contains a finite number of independent modes rather than being a continuum~\cite{PopulationInversion}. In particular, the superohmic baths used in the main text, $J_s^B$, can be well described by a single effective mode~\cite{PopulationInversion}, which we use to obtain accurate analytic expressions for the light transition rates. To best represent the continuum coupling rates, the coupling strength $\check{g}$ and energy $\check{\omega}$ of the single mode must be chosen so that
\begin{equation}
    \check{S}_s\equiv\left|\frac{\check{g}_s}{\check{\omega}_s}\right|^2=j_s^{(0)}, \quad\mathrm{and}\quad
    \check{\omega}_s=\frac{j_s^{(1)}}{j_s^{(0)}},
\end{equation}
where $\check{S}_s$ is the Huang-Rhys parameter of the mode and
\begin{equation}
    j_s^{(i)}=\int_0^{\infty}\mathrm{d}\omega\ \frac{J_s^B(\omega)}{\omega^2}\omega^i,
\end{equation}
are the weighted moments of the spectral density. The spectral density of the effective single mode is therefore $\check{J}^B_s(\omega)=\check{S}_s\check{\omega}_s^2\delta(\omega-\check{\omega}_s)$ and, using this, we find
\begin{equation}
    \mathcal{K}_{ss'}(\omega)\big\vert_{s=s'}=\pi \sum_{p=-\infty}^{\infty}A_{s,p}\delta(\omega-p\check{\omega}_s),
\end{equation}
where
\begin{equation}
    A_{s,p}=e^{-2\check{S}_s\check{N}_s}\sideset{}{'}\sum_{n=\left|p\right|}^{\infty}\sum_{m=\frac{n-p}{2}}^n \check{W}_{s,n}\check{N}_s^m\binom{n}{m}\binom{m}{m-\frac{1}{2}(n-p)},
\end{equation}
$\check{W}_{s,n}=(\check{S}_s^n/n!)\exp(-\check{S}_s)$, $\check{N}_s$ is the thermal population of the mode at temperature $T^B$, and the prime on the summation indicates that only every other term in the sum is included~\cite{PopulationInversion}. 

Using the single-mode expressions for $\mathcal{K}_{ss'}$ and the frequency-independence of $G_{ss'}$, Eq.~\eqref{eq:Iss} becomes
\begin{align}
    M^L_{ss'}(\omega)\big\vert_{s\neq s'}&=\pi\kappa_s\kappa_{s'}G_{ss'}\left[J^A(-\omega)+J^E(\omega)\right],\\
    M^L_{ss'}(\omega)\big\vert_{s= s'}&=\pi G_{ss}\sum_{p=-\infty}^{\infty}A_{s,p}\left[J^A(-\omega+p\check{\omega}_s)+J^E(\omega-p\check{\omega}_s)\right],
\end{align}
where $J^E(\nu)=J^L(\nu)(1+N^L(\nu))$ and $J^A(\nu)=J^L(\nu)N^L(\nu)$ are the thermal spectral densities of the light.

We can now derive $\mathcal{L}^L$ in Eq.~(20) of the main text by defining $\gamma_{\pm}^{\uparrow}=\gamma^\pm(-\delta_{\pm})$, $\gamma_{\pm}^{\downarrow}=\gamma^\pm(\delta_{\pm})$, $\theta_{\pm}^{\uparrow}=\theta(-\delta_{\pm})$ and $\theta_{\pm}^{\downarrow}=\theta(\delta_{\pm})$, where the light rate functions are
\begin{align}
    \gamma^+(\omega)&\equiv2\Gamma^L_{++}(\omega)\\
    &=2\left[\cos^2(\chi/2)M^L_{RR}(\omega)+\sin^2(\chi/2)M^L_{LL}(\omega)+\sin\chi M^L_{LR}(\omega)\right],\\
    \gamma^-(\omega)&\equiv2\Gamma^L_{--}(\omega)\\
    &=2\left[\sin^2(\chi/2)M^L_{RR}(\omega)+\cos^2(\chi/2)M^L_{LL}(\omega)-\sin\chi M^L_{LR}(\omega)\right],
\end{align}
and the light coherence function is
\begin{align}
    \theta(\omega)&\equiv \Gamma^L_{+-}(\omega)=\Gamma^L_{-+}(\omega)\\
    &=\tfrac{1}{2}\sin\chi[M^L_{LL}(\omega)-M^L_{RR}(\omega)]+\cos\chi M^L_{LR}(\omega).
\end{align}
To arrive at these we used that $M_{LR}^L=M_{RL}^L$ and the explicit expressions for the eigenstate amplitudes $c_a^{(s)}$ in terms of the angle $\chi$, as defined in Supporting Information~1 (SI-1). 

Finally, we make the same assumptions as in the main text: degenerate dipoles ($\chi=\pi/2$), identical phonon baths, and polarization choices leading to $G_{LR}=0$. These imply that $\gamma^+(\omega)=\gamma^-(\omega)=\gamma(\omega)$, and that we arrive at Eqs.~(21) and (23) in the main text, with
\begin{equation}
    \mathcal{F}(\omega)=\pi\sum_{p=-\infty}^{\infty}A_{p}\big[J^A(-\omega+p\check{\omega}_s) +J^E(\omega-p\check{\omega}_s)\big],
\end{equation}
where we have dropped the site index $s$ in $A_{s,p}$ because we are assuming that the baths coupled to both donor sites are identical.


\providecommand{\latin}[1]{#1}
\makeatletter
\providecommand{\doi}
  {\begingroup\let\do\@makeother\dospecials
  \catcode`\{=1 \catcode`\}=2 \doi@aux}
\providecommand{\doi@aux}[1]{\endgroup\texttt{#1}}
\makeatother
\providecommand*\mcitethebibliography{\thebibliography}
\csname @ifundefined\endcsname{endmcitethebibliography}
  {\let\endmcitethebibliography\endthebibliography}{}